\documentclass[12pt]{article}
\usepackage{amsmath,amssymb}

\usepackage{graphicx}
\usepackage{wrapfig}

\usepackage{hyperref}
\usepackage{cite}

\def\beq{\begin{eqnarray}}
\def\eeq{\end{eqnarray}}

\newcommand{\be}{\begin{equation}}
\newcommand{\ee}{\end{equation}}
\newcommand{\bea}{\begin{eqnarray}}
\newcommand{\eea}{\end{eqnarray}}
\newcommand{\bg}{\begin{gather}}

\newcommand{\bseq}{\begin{subequations}}
\newcommand{\eseq}{\end{subequations}}

\renewcommand{\ln}{\mathop{\rm ln}\nolimits}

\def\half{\frac{1}{2}}

\def\be{\begin{eqnarray}}
\def\ee{\end{eqnarray}}
\def\lb{\label}

\begin{document}

\title{\textbf{The Wald entropy and 6d conformal anomaly }}

\vspace{2cm}
\author{ \textbf{  Amin Faraji Astaneh$^{1}$ and  Sergey N. Solodukhin$^2$ }} 
\date{}
\maketitle
\begin{center}
\hspace{-0mm}
  \emph{$^1$ School of Particles and Accelerators,\\ Institute for Research in Fundamental Sciences (IPM),}\\
\emph{ P.O. Box 19395-5531, Tehran, Iran}
 \end{center}
\begin{center}
  \hspace{-0mm}
  \emph{ $^2$ Laboratoire de Math\'ematiques et Physique Th\'eorique  CNRS-UMR
7350, }\\
  \emph{F\'ed\'eration Denis Poisson, Universit\'e Fran\c cois-Rabelais Tours,  }\\
  \emph{Parc de Grandmont, 37200 Tours, France}
\end{center}



\begin{abstract}
\noindent { We analyze the Wald entropy for different forms of the conformal anomaly in six dimensions.
In particular we focus on the anomaly which arises in a holographic calculation of  Henningson and Skenderis.
The various presentations of the anomaly differ by some total derivative terms.  We calculate the corresponding Wald entropy for 
surfaces  which do not have an Abelian $O(2)$ symmetry in the transverse direction although the extrinsic
curvature vanishes.  We demonstrate that for this class of surfaces the Wald entropy is different for different forms of the
conformal anomaly. The difference is due to the total derivative terms present  in the anomaly. 
We analyze the conformal invariance of the Wald entropy for the holographic conformal anomaly and demonstrate that
the violation of the invariance  is due to the contributions of the total derivative terms  in the anomaly. Finally, we make more precise the
general form for the Hung-Myers-Smolkin discrepancy.

}
\end{abstract}

\vskip 1 cm
\noindent
\rule{7.7 cm}{.5 pt}\\
\noindent 
\noindent
\noindent ~~~ {\footnotesize e-mails: faraji@ipm.ir,    Sergey.Solodukhin@lmpt.univ-tours.fr}

\newpage
    \tableofcontents
\pagebreak

\newpage

\section{ Introduction}
More than twenty years ago Wald has introduced a very efficient prescription by which the gravitational entropy
can be computed for a given gravitational action \cite{Wald}, \cite{Wald2} (see also \cite{JM}). The typical gravitational action in question 
is polynomial in the Riemann curvature and its covariant derivatives. 
This prescription associates the entropy with a horizon, a co-dimension two surface with very peculiar properties.
In particular, it is assumed that there exists a time-like Killing vector which becomes null on this surface.
This vector generates an Abelian symmetry in the transverse direction to the surface. As a consequence,
 the extrinsic curvature of the surface is vanishing. 
 
In a related concept of the entanglement entropy one associates some entropy with arbitrary  co-dimension two surface, not necessarily a horizon.
The presence of the Abelian symmetry is thus not guaranteed in this case. Nevertheless, there appears to be a certain algorithm  \cite{Solodukhin:2008dh},
\cite{Fursaev:2013fta}, \cite{Dong}, \cite{Camps}
to associate
some entropy with each individual term, dependent on the curvature, in the  quantum effective action. This algorithm certainly deviates from the
one proposed by Wald  if the surface in question is characterized by a non-trivial extrinsic curvature. However, what happens if the extrinsic curvature
is vanishing while the Abelian symmetry is not present? This would be the situation when the Wald algorithm for the entropy  might be thought to still
work. This was tested by Hung, Myers and Smolkin (HMS) in  \cite{Hung:2011xb} for a certain class of six-dimensional geometries with the properties as above where they
compared the Wald entropy due to the $6d$ conformal anomaly with the holographic entropy computed by using the holographic prescription by Ruy and Takanayagi \cite{Ryu:2006ef}.
They have found a discrepancy between these two calculations. In the recent works   \cite{Astaneh:2014sma}  we have made some attempts to explain this discrepancy, for the  alternative attempts see  \cite{Miao:2014nxa}.

In the present note we deliberately ignore the problem of finding an explanation for the HMS discrepancy. Instead, we make a step back and ask the question which should have been 
answered first: what is the general form for the discrepancy?  The answer to this question is not that trivial as it would seem to be. 
The reason is the following. In \cite{Hung:2011xb} it was considered not the most general form for the conformal anomaly and certainly not the one which actually
appears in a holographic calculation as, for instance,  in the paper of Henningson and Skenderis \cite{Henningson:1998gx}.  The difference between
the two  possible forms is due to the total derivatives.  In this note we show that what the Wald entropy is concerned these total derivatives are important and can not
be disregarded.

This note is organized as follows. In section 2 we present the general form for the trace anomaly in six dimensions. This general form includes the
Euler number, three conformal invariants and a set of total derivative terms. In section 2.2 the holographic conformal anomaly calculated by Henningson and Skenderis is presented in this general form and we give the values of the respective conformal charges as well as the charges that correspond to the total derivatives.
In section 3 we give  the formulas for the Wald entropy due to the 6d conformal anomaly, both in a general form and for the holographic conformal anomaly.
In section 4 we consider the HMS geometries and surfaces and compute the Wald entropy for the holographic anomaly presented in two forms, the one in which  it appears in the holographic calculation and the other which involves the representation in terms of the conformal invariants, and find a mismatch.  In section 5 we explain the mismatch
by taking into account the contribution of the total derivatives. The Wald entropy of the latter is non-vanishing in general. 
In section 6 we discuss the conformal invariance of the corresponding entropy. In section 7 we give a general form for the HMS discrepancy.
We conclude in section 8.

\section{ Conformal anomaly in six dimensions}
\subsection{General form for the anomaly}
In a generic conformal field theory in $d=6$ the trace anomaly, modulo the total derivatives,  is a combination of four different terms 
\be
&&<T>={\cal A}={\cal A}_{BI}+{\cal A}_{C}\, , \ \ \nonumber \\
&&{\cal A}_{BI}= aE_6+b_1I_1+b_2I_2+b_3I_3\, , \ \ \  {\cal A}_{C}=\sum_{k=1}^7d_k C_k\, ,
\lb{3}
\ee
where $E_6$ is the Euler density in $d=6$ and, using notations of \cite{Bastianelli:2000hi}, we have 
\be
&&I_1=W_{\alpha\mu\nu\beta}W^{\mu\sigma\rho\nu}W_{\sigma\ \ \ \rho}^{\ \alpha\beta}\, ,\nonumber \\
&&I_2=W_{\alpha\beta}^{\ \ \mu\nu}W_{\mu\nu}^{\ \ \sigma\rho}W_{\sigma\rho}^{\ \ \alpha\beta}\, , \nonumber \\
&&I_3=W_{\mu\alpha\beta\gamma}\Box W^{\mu\alpha\beta\gamma}+W_{\mu\alpha\beta\gamma} (4R^\mu_\nu-\frac{6}{5}R\delta^\mu_\nu)W^{\nu\alpha\beta\gamma}\, ,
\lb{5}
\ee
where $W_{\alpha\beta\mu\nu}$ is the Weyl tensor. In (\ref{3}) the part ${\cal A}_C$ is due to the total derivatives
\be
&&C_1=B_1\, , C_2=B_2+B_6\, , \, C_3=B_3+B_7\, , \nonumber \\
&&C_4=B_4+B_8\, , \, C_5=B_5+B_9\, , \nonumber \\
&&C_6=\frac{1}{9}B_2-B_4-\frac{1}{5}B_{11}-\frac{3}{2}B_{13}+B_{14}\, ,\nonumber \\
&&C_7=\frac{1}{60}B_2-\frac{3}{4}B_3+\frac{3}{4}B_4+\frac{1}{4}B_5+\frac{1}{12}B_{12}+\frac{1}{2}B_{15}-\frac{1}{4}B_{16}-B_{17}\, ,
\lb{6}
\ee
where
\be
\begin{split}
&B_1=\nabla^4 R\, , \, B_2=(\nabla_\alpha R)^2\, , \, B_3=(\nabla_\alpha B_{\mu\nu})^2\, , \, B_4=\nabla_\alpha B_{\mu\nu}\nabla^\mu B^{\alpha\nu}\,  ,\, B_5=(\nabla_\alpha W_{\mu\nu\rho\sigma})^2\, , \\
&B_6=R\nabla^2 R\, , \, B_7=B_{ab}\nabla^2 B^{ab}\, , \,  B_8=B_{\alpha\beta}\nabla_\mu\nabla^\beta B^{\alpha\mu}\, , \, B_9=W_{\alpha\beta\mu\nu}\nabla^2 W^{\alpha\beta\mu\nu}\, , \, B_{10}=R^3\, , \, \\
&B_{11}=R B_{\mu\nu}^2\, , \, B_{12}=RW_{\mu\nu\rho\sigma}^2\, , \, B_{13}=B_\mu\, ^\nu B_\nu\,^\rho B_\rho\, ^\mu\, , \, B_{14}=B_{\alpha\beta}B_{\mu\nu}W^{\alpha\mu\beta\nu}\, \\
&B_{15}=B_{\alpha\beta}W^{\alpha\mu\nu\rho}W^\beta\,_{\mu\nu\rho}\, , \, B_{16}=W_{\alpha\beta}\,^{\mu\nu}W_{\mu\nu}\,^{\rho\sigma}W_{\rho\sigma}\,^{\alpha\beta}\, , \, B_{17}=W_{\alpha\mu\beta\nu}W^{\alpha\rho\beta\sigma}W^\mu\,_\rho\,^\nu\,_\sigma\, ,
\lb{7}
\end{split}
\ee
and  the tensor $B$ is defined as
\be
B_{\mu\nu}=R_{\mu\nu}-\frac{1}{6}Rg_{\mu\nu}\, .
\lb{8}
\ee
The form (\ref{3}), when the contribution of the total derivative terms $C_k$ is neglected, for the conformal anomaly we shall call the $BI$-form.

\subsection{Holographic conformal anomaly}
In this paper, we are mostly  interested in the holographic conformal anomaly computed in 
\cite{Henningson:1998gx}. It is derived by a standard holographic procedure from  the 7-dimensional gravitational action 
\be
W_7=-\frac{1}{2\ell^5_p}\int d^7 x\sqrt{g}(R+\frac{30}{L^2})\, .
\lb{W7}
\ee
This anomaly takes the form
\be
{\cal A}^h=-\frac{L^5}{64\ell_p^5}\big(\half RR_{\mu\nu}R^{\mu\nu}-\frac{3}{50}R^3-R_{\mu\nu\rho\sigma}R^{\mu\rho}R^{\nu\sigma}+
\frac{1}{5}R^{\mu\nu}\nabla_\mu\nabla_\nu R-\half R^{\mu\nu}\Box R_{\mu\nu}+\frac{1}{20}R\Box R\big)\, .
\lb{9}
\ee
It can be represented in the form (\ref{3}) as follows
\be
&&{\cal A}^{h}={\cal A}^h_{BI}+{\cal A}^h_C\, , \nonumber \\
&&{\cal A}^h_{BI}=\frac{3L^5}{2^5\,  7!\ell_p^5}(-\frac{35}{2}E_6-1680I_1-420I_2+140I_2)\, , \nonumber \\
&&{\cal A}^h_C=\frac{3L^5}{2^5\, 7!\ell_p^5}(-140C_5+420C_3-504C_4-84C_6+560C_7)\, .
\lb{10}
\ee
So that we find for the charges $b_k$ and $a$,
\be
b_1=-\frac{L^5}{32\ell_p^5}\, , \, \, b_2=-\frac{L^5}{128\ell_p^5}\, , \, \, b_3=\frac{L^5}{384\ell_p^5}\, , \, \, a=-\frac{L^5}{3072\ell_p^5}
\lb{bk}
\ee
in agreement with \cite{Hung:2011xb}  and \cite{deBoer:2009gx}. We also find that
\be
d_3=-\frac{L^5}{384\ell_p^5}\, , \,\, d_4=-\frac{3L^5}{320\ell_p^5}\, , \, \, d_5=-\frac{L^5}{384\ell_p^5}\, , \, \, d_6=-\frac{L^5}{640\ell_p^5}\, , \, \, 
d_7=\frac{L^5}{96\ell_p^5}\, 
\lb{dk}
\ee
and all other $d_k$ vanish.
\section{The Wald entropy}
\subsection{General formula}
Suppose that the gravitational action $W$ is a function of the Riemann curvature and its covariant derivatives (and contains maximum two derivatives). Then the corresponding Wald entropy is computed according to the formula \cite{Wald2}
\be
S_W=-2\pi \int_\Sigma (\frac{\partial W}{\partial R_{\mu\nu\rho\sigma}}-\nabla_\alpha \frac{\partial W}{\partial \nabla_\alpha R_{\mu\nu\rho\sigma}}+
\nabla_{(\alpha} \nabla_{\beta )} \frac{\partial W}{\partial \nabla_{(\alpha}\nabla_{\beta )} R_{\mu\nu\rho\sigma}})\epsilon_{\mu\nu}\epsilon_{\rho\sigma}\, ,
\lb{1-1}
\ee
where $\epsilon_{\mu\nu}=n^a_\mu n^b_\nu \epsilon_{ab}$, $n^a_\mu$, $a=1,2$ is a pair vectors normal to surface $\Sigma$, so that we have that
\be
\epsilon_{\mu\nu}\epsilon_{\rho\sigma}=(n^a_\mu n^a_\rho)(n^b_\nu n^b_\sigma)-(n^a_\mu n^a_\sigma)(n^b_\nu n^b_\rho)\, .
\lb{2}
\ee
We notice the symmetrization in the last term in (\ref{1-1}). It will be important when we discuss the Wald entropy due to the total derivatives.

\subsection{Entropy due to the holographic conformal anomaly}
Using the  above definition  we compute the corresponding  Wald's  entropy.
In the case of the holographic conformal anomaly (\ref{9}) we find
\be
&&S_W^h=\frac{\pi L^5}{32\ell_p^5}\int_\Sigma [   (RR_{aa}+R_{\mu\nu}^2)-\frac{9}{25} R^2-(2R_{\mu a \nu a}R^{\mu\nu}+R_{aa}^2-R_{ab}^2)\nonumber \\
&&+\frac{1}{5}\nabla_a\nabla_a R- \Box {R}_{aa}+\frac{2}{5}\Box R]\, ,
\lb{11}
\ee
where the indexes $a$ and $b$ correspond to projections on the normal vectors $n^\mu_a$, $a=1,\,  2$ and we introduced the  notations $\nabla_a\equiv n^\mu_a\nabla_\mu$, $\Box {R}_{aa}\equiv n^\mu_a n^\nu_a \Box R_{\mu\nu}$.

\subsection{Entropy due to conformal anomaly in the $BI$-form }
On the other hand, for a generic anomaly in the form ${\cal A}_{BI}$ (\ref{3}) we have that
\be
S_{BI}=(a E_4+b_1s_1+b_2s_2+b_3 s_3)\, ,
\lb{12}
\ee
\be
&&s_1=-6\pi(W^{b\mu\nu a}W_\mu\,^{ab}\,_\nu-W^{a\mu\nu a}W_\mu\,^{bb}\,_\nu-\frac{1}{4}W^{a\mu\nu \sigma}W^a\,_{\mu\nu \sigma}+\frac{1}{20}W^{\mu\nu \sigma\rho}W_{\mu\nu \sigma\rho})\, , \label{13} \\
&&s_2=-6\pi(2W^{ab\mu\nu}W_{\mu\nu}\,^{ab}-W^{a\mu\nu \sigma}W^a\,_{\mu\nu \sigma}+\frac{1}{5}W^{\mu\nu  \sigma\rho}W_{\mu\nu \sigma\rho})\, , \nonumber \\
&&s_3=-8\pi(\Box W^{abab}+4R^a_\mu W^{\mu bab}-R_{\mu\nu }W^{\mu a\nu a}-\frac{6}{5}RW^{abab}+W^a\,_{\mu\nu \sigma}W^{a\mu\nu\sigma}-\frac{3}{5}W_{\mu\nu\sigma\rho}W^{\mu\nu\sigma\rho})\, ,\nonumber
\ee
as  was first derived in \cite{Hung:2011xb}. Notice that the contribution of the total derivative terms $C_k$ is neglected in (\ref{12}).  This contribution is indeed supposed to vanish when considered in the standard situation of a Killing horizon. 

\section{The Wald entropy for the HMS-surfaces}

\subsection{The HMS geometries/surfaces}
Hung, Myers and Smolkin \cite{Hung:2011xb} have considered the following six-dimensional geometries and the four-dimensional surfaces:
\begin{itemize}
\item
\textbf{a)}\ $\Sigma: S^1\times S^3$ in $R\times S^2\times S^3$, $R_1$ is radius of $S^3$, $R_2$ is radius of $S^2$\, ,
\item
\textbf{a')}\ $\Sigma: S^2\times S^2$ in $R\times S^2\times S^3$, $R_1$ is radius of $S^3$, $R_2$ is radius of $S^2$\, ,
\item
\textbf{b)}\ $\Sigma: R^2\times S^2$ in $R^3\times S^3$, $R_1$ is radius of $S^3$\, ,
\item
\textbf{c)}\ $\Sigma: R^1\times S^3$ in $R^2\times S^4$,  $R_1$ is radius of $S^4$\, .
\end{itemize}
All these geometries are characterized by same properties:

\medskip

\noindent i) they are products of constant curvature spaces; that is why, in the Wald entropy (\ref{11}) or (\ref{13})  all terms  with covariant derivatives  vanish;

\medskip

\noindent ii) the time coordinate $x^1=t$ lies in the flat sub-space, so that components
of the Riemann tensor $R_{\mu\nu\alpha\beta}$ or the Ricci tensor $R_{\mu\nu}$ vanish if one of the indexes is $1$;

\medskip

\noindent iii) the surface $\Sigma$ has two normal vectors, one of which, $n^1$,  is  time-like. The corresponding extrinsic curvature vanishes because
it is the Killing vector.  The surface $\Sigma$ has a component which is the minimal surface embedded in a sphere. The corresponding normal vector, $n^2$, lies along this
sphere. The corresponding extrinsic curvature vanishes since  it is the minimal surface;

\noindent vi) in particular, we have that $R^2_{aa}=R^2_{ab}$ and $R_{abcb}=0$ for all these geometries; 
 
 \medskip

\noindent v) in all these cases the Abelian $O(2)$ symmetry in the transverse direction to $\Sigma$ is absent although the components of the extrinsic curvature of $\Sigma$
vanish.

\subsection{Entropy due to holographic conformal anomaly}
First, we compute the entropy (\ref{11}) due to the holographic conformal anomaly:
\be
&&{\bf a).} \ \ S_W^h=\frac{\pi}{400}\frac{l^5}{l_p^5} V_\Sigma(\frac{7}{R_2^4}-\frac{12}{R_1^4}-\frac{33}{R_1^2R_2^2})\, ,
\lb{a2}\\
&&{\bf a').}  \ \ S_{W}^h=\frac{\pi}{400}\frac{l^5}{l_p^5} V_\Sigma(\frac{7}{R_2^4}+\frac{38}{R_1^4}-\frac{58}{R_1^2R_2^2})\, ,
\lb{a3} \\
&&{\bf b).} \ \ S_{W}^h=\frac{19\pi}{200}\frac{l^5}{l_p^5} \frac{V_\Sigma}{R_1^4}\, ,
\lb{a4}\\
&&{\bf c).} \ \ S_{W}^h=\frac{27\pi}{400}\frac{l^5}{l_p^5} \frac{V_\Sigma}{R_1^4}\, ,
\lb{a5}
\ee
where $V_\Sigma$ is volume of $\Sigma$.

\subsection{Entropy due to  holographic conformal anomaly in $BI$-form}
Now we compute the Wald entropy for the same holographic anomaly but represented in the $BI$ form (\ref{10}).
We find
\be
&&{\bf a).} \ \ S^{h}_{BI}=\frac{\pi}{400}\frac{l^5}{l_p^5} V_\Sigma(\frac{7}{R_2^4}-\frac{12}{R_1^4}-\frac{33}{R_1^2R_2^2})\, , \lb{b1}\\
&&{\bf a').}  \ \ S^{h}_{BI}=\frac{\pi}{400}\frac{l^5}{l_p^5} V_\Sigma(\frac{7}{R_2^4}+\frac{64}{3R_1^4}-\frac{58}{R_1^2R_2^2})\, , \lb{b2}\\
&&{\bf b).} \ \ S^{h}_{BI}=\frac{4\pi}{75}\frac{l^5}{l_p^5} \frac{V_\Sigma}{R_1^4}\, , \lb{b3}\\
&&{\bf c).} \ \ S^{h}_{BI}=-\frac{23\pi}{400}\frac{l^5}{l_p^5} \frac{V_\Sigma}{R_1^4}\, . \lb{b4}
\ee
These results for the Wald entropy agree with the entropy computed in \cite{Hung:2011xb} when we choose the particular values (\ref{bk}) for the  conformal charges $b_k$.

\subsection{The mismatch}
Clearly, we have a mismatch, $\Delta S^h= S^h_W-S^h_{BI}$, between these two calculations:
\be
&&{\bf a).} \ \ \Delta S^h=0\, , \lb{c1}\\
&&{\bf a').}  \ \ \Delta S^{h}=\frac{\pi}{24}\frac{l^5}{l_p^5} \frac{V_\Sigma}{R_1^4}\, , \lb{c2}\\
&&{\bf b).} \ \  \Delta S^{h}=\frac{\pi}{24}\frac{l^5}{l_p^5} \frac{V_\Sigma}{R_1^4}\, , \lb{c3}\\
&&{\bf c).} \ \ \Delta S^{h}=\frac{\pi}{8}\frac{l^5}{l_p^5} \frac{V_\Sigma}{R_1^4}\, . \lb{c4}
\ee

\section{Contribution of the total derivative terms}
\subsection{Some general formulas}
Trying to understand the mismatch which we have just observed we have to consider carefully the total derivative terms which are in general present
in the conformal anomaly. As wee see from (\ref{10}) only terms $C_k\, , \, \, k=3,4,5,6,7$ contribute to the holographic conformal anomaly.
Before using the Wald formula (\ref{1-1}) and compute the entropy it is required to first transform each $C_k$ to a form which would contain 
terms symmetric in second covariant derivatives of the curvature. The terms $C_3$ and $C_5$ already take this form. The corresponding Wald entropy
calculated using (\ref{1-1}) gives zero entropy,
\be
S_W^{C_3}=S_W^{C_5}=0\, .
\lb{21}
\ee 
For the other terms the situation is less trivial.
Commuting the covariant derivatives  one finds
\be
&&C_4=D_4+R_\alpha\,^\beta R_\beta\,^\mu R_\mu\,^\alpha -R^{\alpha\beta}R^{\mu\nu}R_{\alpha\mu\beta\nu}\, \nonumber \\
&&C_7=D_7+I_1-\frac{1}{4}I_2+\frac{1}{2}R_{\alpha\beta}W^{\alpha\mu\nu\rho}W^\beta\,_{\mu\nu\rho}\, , \nonumber \\
&&C_6=D_6+R_{\mu\nu}R_{\alpha\beta}R^{\mu\alpha\nu\beta}-R_\mu\,^\alpha R_\alpha\,^\beta R_\beta\,^\mu\, ,
\lb{C}
\ee
where $I_1$ and $I_2$ are defined in (\ref{2})  and we introduced
\be
&&D_4=-\frac{1}{18}R\Box R-\frac{5}{36}(\nabla_\alpha R)^2+\frac{1}{3}R^{\alpha\beta}\nabla_\alpha \nabla_\beta R+\nabla_\alpha R_{\mu\nu}\nabla^\nu R^{\alpha\mu}\, ,\nonumber \\
&&D_6=\frac{1}{4}(\nabla_\alpha R)^2-\nabla_\mu R^{\alpha\beta}\nabla_\beta R^\mu\,_\alpha\, ,\nonumber \\
&&D_7=\frac{1}{4}(\nabla_\alpha R_{\mu\nu\rho\sigma})^2-(\nabla_\alpha R_{\mu\nu})^2+\frac{1}{16}(\nabla_\alpha R)^2+\frac{3}{4}\nabla_\mu R^{\alpha\beta}\nabla_\beta R^\mu\,_\alpha\, .
\lb{D}
\ee
Now we apply the general formula (\ref{1-1}) and compute  the Wald entropy,
\be
&&S^{(C_4)}_W=2\pi \int_\Sigma [-\frac{2}{3}(\Box-\nabla_a^2)R+(R^2_{aa}-R^2_{ab}-R^2_{\mu a})]\, , \lb{SS} \\
&&S^{(C_6)}_W=2\pi \int_\Sigma [(\Box-\nabla^2_a)R-(R^2_{aa}-R^2_{ab}-R^2_{\mu a})]\, ,\nonumber \\
&&S^{(C_7)}_W=s_1-\frac{1}{4}s_2+2\pi \int_\Sigma [ \Box(R_{abab}-2R_{aa}+\frac{1}{4}R)+\frac{3}{4}\nabla_a^2 R +\frac{3}{2}(R^2_{\mu a}-R_{\mu\nu}R^{\mu a\nu a})\nonumber \\
&&-2\pi\int_\Sigma (\half W^{a\mu\nu\rho}W^a\,_{\mu\nu\rho}+2R^a_\mu W^{\mu bab}-\frac{1}{2}R_{\mu\nu}W^{\mu a \nu a})\, ,\nonumber
\ee
where we define $\Box R_{abab}=n_a^\mu n^\nu_a n^\alpha_b n^\beta_b \Box R_{\mu\alpha\nu\beta}$ and $\Box R_{aa}=n^\mu_a n^\nu_a\Box R_{\mu\nu}$, and the terms  $s_1$ and $s_2$ have been defined in (\ref{13}).  We used the Bianchi identities and, in particular,   that
\be
\nabla_\alpha\nabla_\nu R^\alpha_{\ \mu}-\frac{1}{2}\nabla_\nu\nabla_\mu R=R_{\alpha\mu}R^\alpha_{\ \nu}-R^\alpha_{\ \mu\beta\nu}R^\beta_{\ \alpha}
\lb{rel}
\ee
when derived (\ref{SS}). This identity, in particular,  shows that the right hand side of (\ref{rel}), and its projections on the transverse subspace,  vanishes for a product of constant curvature spaces.  We then use that if the extrinsic curvature of surface $\Sigma$ vanishes we have the following relation
\be
(\Box-\nabla_a^2)R=\Delta_\Sigma R \, ,
\lb{rel2}
\ee
where $\Delta_\Sigma$ is the Laplace operator defined on surface $\Sigma$ and $R_\Sigma$ is the intrinsic curvature of $\Sigma$. 
The integration of  (\ref{rel2}) over a closed surface $\Sigma$  gives zero and the expressions (\ref{SS}) are simplified
\be
&&S^{(C_4)}_W=-S^{(C_6)}_W=2\pi \int_\Sigma (R^2_{aa}-R^2_{ab}-R^2_{\mu a})\, , \lb{SS2}  \\
&&S^{(C_7)}_W=s_1-\frac{1}{4}s_2+\pi \int_\Sigma [2\Box (R_{abab}-2R_{aa}+R)  +3(R^2_{\mu a}-R_{\mu\nu}R^{\mu a\nu a})\nonumber \\
&&-W^{a\mu\nu\rho}W^a\,_{\mu\nu\rho}-4R^a_\mu W^{\mu bab}+R_{\mu\nu}W^{\mu a \nu a}]\, .
\nonumber
\ee
We note that these formulas are valid for arbitrary geometry and any entangling surface for which the extrinsic curvature vanishes.
If the geometry in question is a product of constant curvature spaces, as in the examples considered in section 4.1,  then the terms with 
derivatives in (\ref{SS2}) vanish. 
Each entropy in  (\ref{SS2}) is a priori non-vanishing and we shall demonstrate this below for the concrete examples. 
Accidentally, we observe that the Wald entropy for invariants $C_4$ and $C_6$ is the same,  up to a sign.

\subsection{Explaining the mismatch}
Now let us compute the contribution of these to total derivative terms to the entropy, for various geometries discussed before. Our aim is to demonstrate that
\be
\Delta S^h=(d_4S_W^{(C_4)}+d_6S_W^{(C_6)}+d_7S_W^{(C_7)})\, .
\lb{Sm}
\ee
 The values for the charges $d_k$ are given in (\ref{dk}).
 
 \bigskip

We find in the following cases:

\bigskip

\noindent \textbf{a)}\ $\Sigma: S^1\times S^3$ in $R\times S^2\times S^3$\, :

\be
S_W^{(C_4)}=-S_W^{(C_6)}=-\frac{2\pi V_\Sigma L^5}{\ell_p^5 R_2^4}\, , \  \ S_W^{(C_7)}=-\frac{3\pi V_\Sigma L^5}{2\ell_p^5R_2^4}\, , 
\ee
and hence
\be
\sum_k d_k S_W^{(C_k)}=0\, .
\ee

\noindent\textbf{a')}\ $\Sigma: S^2\times S^2$ in $R\times S^2\times S^3$\,:

\be
S_W^{(C_4)}=-S_W^{(C_6)}=-\frac{8\pi V_\Sigma L^5}{\ell_p^5R_1^4}\ , \  \ S_W^{(C_7)}=-\frac{2\pi V_\Sigma L^5}{\ell_p^5R_1^4}\, ,
\ee
and hence
\be
\sum_k d_k S_W^{(C_k)}=\frac{\pi V_\Sigma L^5}{24\ell_p^5R_1^4} \, .
\ee

\noindent\textbf{b)}\ $\Sigma: R^2\times S^2$ in $R^3\times S^3$\,:

\be
S_W^{(C_4)}=-S_W^{(C_6)}=-\frac{8\pi V_\Sigma L^5}{\ell_p^5R_1^4} \ , \  \ S_W^{(C_7)}=-\frac{2\pi V_\Sigma L^5}{\ell_p^5R_1^4} \, , 
\ee
and hence
\be
\sum_k d_k S_W^{(C_k)}=\frac{\pi V_\Sigma L^5}{24\ell_p^5R_1^4} \, .
\ee

\noindent\textbf{c)}\ $\Sigma: R^1\times S^3$ in $R^2\times S^4$\, :

\be
S_W^{(C_4)}=-S_W^{(C_6)}=-\frac{18\pi V_\Sigma L^5}{\ell_p^5R_1^4} \ , \  \ S_W^{(C_7)}=-\frac{3\pi V_\Sigma L^5}{2\ell_p^5R_1^4} \, ,
\ee
and hence
\be
\sum_k d_k S_W^{(C_k)}=\frac{\pi V_\Sigma L^5}{8\ell_p^5R_1^4} \, .
\ee
We see that in all these cases we obtain a complete agreement with (\ref{c1})-(\ref{c4}) and  confirm (\ref{Sm}). The mismatch thus is indeed due to the contributions of the total derivative terms which appear
in the conformal anomaly when we pass from the form (\ref{9})   to the $BI$-form  (\ref{10}).

\section{Conformal invariance}
In this section we would like to check whether the Wald entropy in question is conformal invariant.
We restrict ourselves to the infinitesimal conformal transformations, $\delta_\sigma g_{\mu\nu}=g_{\mu\nu}\sigma$. Then we find in six dimensions that
\be
&&\delta_\sigma R=-5\Box \sigma-R\sigma\, , \, \, \delta_\sigma R_{\mu\nu}=-2\nabla_\mu\nabla_\nu \sigma -\frac{1}{2}g_{\mu\nu}\Box \sigma\, ,\, \,\delta_\sigma R_{aa}=-2\nabla^2_a\sigma-\Box \sigma-\sigma R\, , \nonumber \\
&&\delta_\sigma R_{\mu a\nu a}=-\frac{1}{2}(g_{\mu\nu}\nabla^2_a\sigma-n_\nu^a\nabla_a\nabla_\mu\sigma-n_\mu^a\nabla_a\nabla_\nu\sigma+2\nabla_\mu\nabla_\nu \sigma)\, .
\lb{CC}
\ee
First, we analyze the conformal invariance of the Wald entropy (\ref{11}) due to  the holographic conformal anomaly.  In order to simplify the transformations
we consider the case of a geometry with constant curvature, i.e. the condition $\nabla_\rho R_{\alpha\beta\mu\nu}=0$ is assumed to be valid. 
Then we find for the variations of the quantities, integrated over the surface $\Sigma$,
\be
&&\delta_\sigma \int_\Sigma \Box R=-\int_\Sigma (5\Box^2\sigma +R\Box \sigma)\, , \lb{CC1} \\
&&\delta_\sigma \int_\Sigma \nabla^2_a R=-\int_\Sigma (5\nabla^2_a \Box \sigma+R\nabla_a^2\sigma)\, , \nonumber \\
&&\delta_\sigma \int_\Sigma \Box R_{aa}=-\int_\Sigma (2\Box\nabla_a^2\sigma +\Box^2\sigma +R_{aa}\Box \sigma)\, , \nonumber \\
&&\delta_\sigma \int_\Sigma \Box R_{abab}=-\int_\Sigma (R_{abab}\Box\sigma +\Box\nabla_a^2\sigma)\, .\nonumber 
\ee
Since the extrinsic curvature of $\Sigma$ vanishes we have $\Box \sigma=\nabla^2_a\sigma+\Delta_\Sigma\sigma$. Finally, we use the commutation relation
\be
(\Box \nabla_a^2-\nabla^2\Box)\sigma=2(R_{ab}-R_{acbc})\nabla_a\nabla_b\sigma\, , 
\lb{Com}
\ee
where $\nabla_a=n_a^\mu\nabla_\mu$, $a=1,\, 2$ are derivatives in the transverse sub-space and we neglected all terms with derivatives along the surface $\Sigma$ since, after integration over $\Sigma$, these terms
give zero. 
Putting everything together we find
\be
\delta_\sigma S^h_W=-\frac{\pi L^5}{8\ell_p^5}\int_\Sigma [(R_{ab}-\frac{1}{2}R_{cc}\delta_{ab})\nabla_a\nabla_b\sigma]\, 
\lb{CI2}
\ee
for the conformal transformation of the Wald entropy (\ref{11}) due to the holographic conformal anomaly.
This conformal transformation vanishes in many particular cases. For instance, it vanishes if the six-dimensional  space-time is Einstein, i.e. $R_{\mu\nu}=\Lambda g_{\mu\nu}$.
In the presence of $O(2)$ symmetry in the transverse sub-space we have that $R_{ab}=\lambda \delta_{ab}\, , \, a=1,\, 2$ and the tensor $R_{ab}-\frac{1}{2}R_{cc}\delta_{ab}=0$ so that
(\ref{CI2}) vanishes, as expected.  However, for the geometries  considered in section 4.1 we have that $R_{11}=0$ and $R_{22}\neq 0$. In this case the right hand side 
of (\ref{CI2}) is non-zero.  It is proportional to $R_{22} (\nabla_2^2-\nabla^2_1)\sigma$. This quantity is non-vanishing if $\sigma$  is a generic function of coordinates $x^1$ and $x^2$ orthogonal to surface $\Sigma$. 

The terms (\ref{13}) in the Wald entropy that are due to the conformal invariants $I_1$, $I_2$  are conformal invariant by construction. Conformal invariance of entropy $s_3$ which is due to invariant $I_3$ is less obvious since $s_3$ is expressed  not only in terms of the Weyl tensor but also in terms of the Ricci tensor, Ricci scalar and a Laplacian. On a background of a constant curvature geometry we find for the conformal variation of $s_3$,
\be
\delta_\sigma s_3=-24\pi\int_\Sigma (\Box\sigma W_{abab}-2\nabla_a\nabla_b\sigma W_{acbc})=-24\pi \int_\Sigma (\delta_{ab}\nabla^2_d\sigma-2\nabla_a\nabla_b\sigma)R_{acbc}\, ,
\lb{X}
\ee 
where in the second equality we have used  that $\Box\sigma=\Delta_\Sigma+\nabla_d^2$  provided the extrinsic curvature of $\Sigma$ vanishes
and  that 
\be
W_{acbc}=R_{acbc}+\delta_{ab}(\frac{1}{20}R-\frac{1}{4}R_{cc})\, .
\lb{X1}
\ee
Now, irrespectively of the presence or absence of the $O(2)$ symmetry we always have that
\be
R_{acbc}=\frac{1}{2}\delta_{ab}R_{dcdc}\, .
\lb{X2}
\ee
Hence, the transformation (\ref{X}) vanishes and the Wald entropy $s_3$ is indeed conformal invariant to linear order  for a constant curvature geometry and a surface with vanishing 
extrinsic curvature. Notice, that for the HMS geometries we have that $R_{acbc}=0$.

That the total  Wald  entropy (\ref{11}) is not conformal invariant is thus due to the presence of the terms (\ref{SS})  which correspond to the total derivative terms in the conformal anomaly.
Indeed, we find for the conformal transformation of these terms
\be
&&\delta_\sigma S^{(C_4)}_W=-\delta_\sigma S^{(C_6)}_W =16\pi\int_\Sigma [  (R_{ab}-\frac{1}{2}R_{cc}\delta_{ab})\nabla_a\nabla_b\sigma]\, , \nonumber \\
&&\delta_\sigma S^{(C_7)}_W=0\, ,
\lb{CI3}
\ee
where in the last line we used (\ref{CC1}) and  (\ref{X1}). Thus, to this order the entropy $S^{(C_7)}_W$ is conformal invariant while the non-invariance of the Wald entropy (\ref{11}) is entirely due to the terms $S^{(C_4)}_W$ and $S^{(C_6)}_W$ (provided we use values (\ref{dk}) for the charges $d_k$).

\section{The HMS discrepancy for a general 6d  anomaly}
The discrepancy discovered in \cite{Hung:2011xb} is a mismatch between the holographic calculation of the universal term in the entropy and the Wald entropy
computed for the holographic conformal anomaly. However, in \cite{Hung:2011xb} it was considered only the $BI$-form of the conformal anomaly and  the total derivative terms were neglected.
As we explained in this note the actual holographic anomaly differs from the $BI$-form by certain total derivative terms. For the geometries considered in \cite{Hung:2011xb}
the Wald entropy of these total derivatives is non-vanishing, as we have just demonstrated. Thus, the finding of \cite{Hung:2011xb} should be completed by adding the corresponding
contributions of the total derivative terms.

For an arbitrary conformal anomaly taking the form (\ref{3})  with the conformal charges $b_k$ and the charges $d_k$ the discrepancy takes the form 
\be
&&\Delta S_{EE}=4\pi b_3 \ln\epsilon \int_\Sigma (W^{ab\mu\nu}W_{\mu\nu}\,^{ab}-W^{a\mu\nu\rho}W^a\,_{\mu\nu\rho}+2W^{a\mu\nu a}W_\mu\,^{bb}\,_\nu-2W^{b\mu\nu a}W_\mu\,^{ab}\,_\nu)\nonumber \\
&&-2\pi(d_4-d_6) \ln\epsilon\int_\Sigma (R^2_{aa}-R^2_{ab}-R^2_{\mu a})   \lb{AS}\\
&&-2\pi d_7 \ln\epsilon \int_\Sigma[2\Box (R_{abab}-2R_{aa}+R)  +3(R^2_{\mu a}-R_{\mu\nu}R^{\mu a\nu a})\nonumber \\
&& +3W^{b\mu\nu a}W_\mu\,^{ab}\,_\nu-3W^{a\mu\nu a}W_\mu\,^{bb}\,_\nu-\frac{3}{2}W^{ab\mu\nu}W_{\mu\nu}\,^{ab}
+\half W^{a\mu\nu\rho}W^a\,_{\mu\nu\rho}+2R^a_\mu W^{\mu bab}-R_{\mu\nu}W^{\mu a \nu a}]\, ,\nonumber 
\ee
where in the first line we used the result of \cite{Hung:2011xb}. This is a general form for the discrepancy. For the geometries more general than those considered in section 4.1
there may appear some terms with derivatives of the curvature in the first line of (\ref{AS}). The analysis of \cite{Hung:2011xb} does not allow us to identify those terms.
We also notice that $d_4-d_6=b_2$ for the holographic anomaly obtained from the $d=7$  Einstein action.

\section{Conclusions}
In this note we have analyzed the general form for the conformal anomaly in six dimensions and the corresponding Wald entropy.
We have demonstrated that what the Wald entropy is concerned the total derivative terms which generically appear in the conformal anomaly 
can not be neglected. Their contribution is essential for the consistency when the entropy of the holographic conformal anomaly is compared to 
the  entropy computed for  the anomaly expressed in terms of the conformal invariants. 
The other important role of the total derivative terms is that their presence in the anomaly breaks the conformal invariance of the corresponding 
Wald entropy. Finally, we use our findings and give a general form for the discrepancy found in \cite{Hung:2011xb}.

\section*{Acknowledgements} 
A.F.A would like to acknowledge the support by  Iran National Science Foundation (INSF).

\end{document}